\documentclass{article}
\usepackage{amsmath}
\usepackage{amsfonts}

\usepackage{amsmath,amssymb}
\usepackage{graphicx}
\usepackage{color}

\newtheorem{theorem}{Theorem}
\newtheorem{lemma}{Lemma}
\newtheorem{corollary}{Corollary}
\newtheorem{property}{Property}
\newtheorem{definition}{Definition}
\newtheorem{proposition}{Proposition}
\newtheorem{remark}{Remark}

\newtheorem{example}{Example}

\newcommand{\F}{\ensuremath{\mathbb F}}
\newcommand{\Z}{\ensuremath{\mathbb Z}}

\newcommand{\done}{\hfill $\Box$ }

\newcommand{\notequiv}{{\,\not\equiv\, }}

\newcommand{\ls}[1]
    {\dimen0=\fontdimen6\the\font\lineskip=#1\dimen0
     \advance\lineskip.5\fontdimen5\the\font
     \advance\lineskip-\dimen0
     \lineskiplimit=0.9\lineskip
     \baselineskip=\lineskip
     \advance\baselineskip\dimen0
     \normallineskip\lineskip\normallineskiplimit\lineskiplimit
     \normalbaselineskip\baselineskip
     \ignorespaces}

\begin{document}

\bibliographystyle{abbrv}

\title{New Sequences Design from Weil Representation with Low Two-Dimensional Correlation in Both Time and Phase Shifts}
\author{Zilong Wang$^{*1}$ and Guang Gong$^2$ \\
$^1$ School of Telecommunication Engineering, Xidian University, \\
Xi'an, 710071,  P.R.CHINA\\
$^2$ Department of Electrical and Computer Engineering, University of Waterloo \\
Waterloo, Ontario N2L 3G1, CANADA \\
Email: wzlmath@gmail.com \ \ \ \ ggong@calliope.uwaterloo.ca\\
}

%\date{}
 \maketitle

\footnotetext[0] {$^*$ The work was conducted when Zilong Wang was a visiting Ph.D
student at the Department of ECE in University of Waterloo from
September 2008 to August 2009.}

\thispagestyle{plain} \setcounter{page}{1}

\begin{abstract}

A new elementary expression  of the construction first proposed by Gurevich, Hadani, and Sochen is
given, which avoids the explicit use of the Weil representation.
The sequences  in this signal set are
given by both multiplicative character and additive
character of finite field $\mathbb{F}_p$.  Such a signal set
consists of $p^2(p-2)$  time-shift distinct sequences, the magnitude
of the two-dimensional autocorrelation function (i.e., the ambiguity
function) in both time and phase of each sequence is upper bounded
by $2\sqrt{p}$ at any shift not equal to $(0, 0)$.  Furthermore, the
magnitude of their Fourier transform spectrum is less than or equal
to $2$.  For  a subset  consisting of  $p(p-2)$ phase-shift distinct sequences in this signal set,   the magnitude
of the ambiguity function of any pair is upper bounded  by $4\sqrt{p}$.  A proof is given through finding a new expression
of the sequences in the finite harmonic oscillator system. An open problem for
directly establishing these assertions without involving the Weil
representation is addressed.


\end{abstract}

\ls{1.5}
\section{Introduction}

Sequence design for good correlation finds many important
applications in various transmission systems in communication
networks, and radar systems.

\vspace{0.1in} \noindent {\em A.  Low Correlation}

In code division multiple access (CDMA) applications of spread
spectral communication, multiple users share a common channel. Each
user is assigned a different spreading sequence (or spread code) for
transmission. At an intended receiver, despreading (recovering the
original data) is accomplished by the correlation of the received
spread signal with a synchronized replica of the spreading sequence
used to spread the information where the spreading sequences used by
other users are treated as interference, which is referred to as
{\em multiple access interference}. This type of interference, which
is different from interference that arises in radio-frequency (RF)
communication channels, can be reduced by proper design of a
spreading signal set. The performance of a signal (or sequence) set
used in a CDMA system is measured by the parameters $L$, the length
or period of a sequence in the set, $r$, the number of
time-shift distinct sequences, and $\rho$, the maximum magnitude of
the out-of-phase autocorrelation of any sequence and cross
correlation of any pair of the sequences in the set. This is
referred to as  an $(L, r, \rho)$ {\em signal set}. The trade-off of
these three parameters is bounded by the Welch bound, established
in 1974 by Welch \cite{Welch}. The research for constructing good
signal sets has flourished in the literature. The reader is
referred to \cite{Chu72, Alltop80, Suehiro88, Pop92, Brodzik09, FanDarnell} for polyphase sequences with  large
alphabet sizes,  \cite{KumarMoreno90, Hammons-award94} for
$\Z_4$ sequences, \cite{Gong95, Paterson98} for interleaved
sequences, and \cite{SarwartePursley, KumarHelleseth-chapter98, GolombGong-book05} in general,
for example.

\vspace{0.1in} \noindent {\em B.  Minimized Fourier Spectrum}

The orthogonal frequency division multiplexing (OFDM) utilizes
the concept of parsing the input data into $N$ symbol streams, and each
of which in turn is used to modulate parallel, synchronous
subcarriers. With an OFDM system having $N$ subchannels, the symbol
rate on each subcarrier is reduced by a factor of $N$ relative to
the symbol rate on a single carrier system that employs the entire
bandwidth  and transmits data at the same rate as OFDM. An OFDM
signal can be implemented  by computing an inverse Fourier
transform and  Fourier transform at  the transmitter side and
receiver side, respectively.  A major problem with the multicarrier
modulation in general and OFDM system in particular is the high
peak-to-average power ratio (PAPR) that is inherent in the
transmitted signal. A bound on PAPR
through the magnitude of the discrete Fourier transform (DFT) spectrum of
employed signals is shown in \cite{Litsyn1, PatersonTarokh00}.  (See  \cite{Terras} for  details of  Fourier transform.)  One way to achieve  low PAPR is to employ Golay complementary sequences,
as first shown by Davis and Jedwab in  \cite{DavisJedwab99}.
A tremendous amount of work has been done along this line
since then.

\vspace{0.1in} \noindent {\em C. Low Valued Ambiguity Functions}

In radar or sonar applications, a sequence should be designed in
such a way that the {\em ambiguity function} (the two-dimensional
autocorrelation function in both time and frequency or equivalently
phase, will be formally defined later), having the value of the
length of the sequence at $(0, 0)$, and small values at any shift
not $(0, 0)$. The {\em ambiguity function} is required for
determining the {\em range} (proportional to the time-shift) and
{\em Doppler}  (the velocity to or from the observer, proportional
to the frequency shift) of a target. Sequences with low
ambiguity function can be achieved by Costas arrays, which yield the
so-called {\em ideal} or {\em thumb-tack} ambiguity function (which only takes
the values 0 or 1 at any shift not (0, 0))
\cite{Costas84, GolombGong07}.

It is interesting to see whether there exists a signal set which  simultaneously satisfies the
requirements that arise from the above three transmission scenarios,
i.e., having low correlation, low PAPR, low ambiguity function, but with  large size and moderate implementation cost. It is  anticipated that employing those sequences will improve the
performance of communication systems with multi-carrier CDMA
transmission \cite{Proakis}, radar networks, and transmission
systems in future cognitive radio networks \cite{Sampath407}.

Gurevich, Hadani, and Sochen \cite{S.Gurevich, S.Gurevich3} proposed
a signal set called finite oscillator system $\mathfrak{S}$ which gives a positive answer to the above question except for the implementation cost.
Their construction makes use of
the group-theoretic Weil representation and  the sequences are   described in  algorithmic terms  by  the end of \cite{S.Gurevich, S.Gurevich2}.
The main contribution of this paper is  to propose   a simple  elementary expression for those sequences, which  avoids the need to
explicitly employ relatively costly group-theoretic computations.

It is interesting to observe that to date, almost all sequences with low correlation
in the literature are related to the use of additive or multiplicative characters
of the finite field or Galois rings together with functions. Recently, inspired by
mutually unbiased bases discussed by Howe in \cite{R.Howe}, Howard, Calderbank, and Moran \cite{S.D.Howard} investigated
sequences constructed from the Heisenberg representation in 2006, then Gurevich, Hadani, and Sochen \cite{S.Gurevich, S.Gurevich3} introduced sequences from the Weil representation in 2008, which are
referred to as a finite oscillator system $\mathfrak{S}$.

In fact, sequences from the Heisenberg representation are related to extended  a Frank-Zadoff-Chu (FZC)
sequence \cite{FrankZadoff62, Chu72, Frank73},
being complex valued sequences with period $p$. After normalization
by the energy, the values of their ambiguity functions (precisely defined in the next section) is bounded by
$\frac{1}{\sqrt{p}}$ except for some special
case. While the sequences from the Weil representation, which will
be introduced later, have the desired properties in the above
mentioned three application scenarios, but have a complicated form.  Gurevich, Hadani, and Sochen investigated how to implement
their sequences in terms of an algorithm.  The goal of this paper is to find a simple elementary expression for the finite harmonic oscillator system.    We show that there are two types of the  sequences in the  finite harmonic oscillator system  of the splitting case  (we will formally define it later).   Sequences of the  first type  can be  given as product sequences using both multiplicative characters and additive characters of the
finite field $\F_p$,  and  sequences of  the second type  are  involved the summations of  sequences of the first type.   We construct a new signal set from  the set consisting of the sequences of the first type  with some extension.

The rest of the paper is organized as follows. In Section 2, we
introduce some basic concepts and notations. In
Section 3, we present our new constructions and the main results. In
Section 4, we introduce Weil
representation and the finite oscillator system
constructed by Gurevich, Hadani and Sochen in \cite{S.Gurevich, S.Gurevich3}. We show a simple elementary expression for
this finite oscillator system, and present a proof for the new
 constructions  in Section 5. Comparisons of the new constructions with some known constructions are made in Section 6. Section
7 is for concluding remarks and addressing an open problem.

\section{Basic Concepts and Definitions}

In this section, we introduce some basic concepts and notations
which are frequently used in this paper. For a given prime $p$,  let
$\theta$ and $\eta$  denote the $(p-1)$th and $p$th primitive roots
of unity in the complex field respectively, i.e.,
$$\theta=\exp \left( \frac{2\pi i}{p-1}\right)\ \  \mbox{and} \ \ \eta=\exp\left(\frac{2\pi i}{p}\right).$$

We denote $\mathbb{F}_p$ as the finite field with $p$ elements, and
$\mathbb{F}_p^*$ as the multiplicative group of $\mathbb{F}_p$ with
a generator $\alpha$. Then for every element $\beta\in \mathbb{F}_p^*$, there
exists $i$ with $0\leqslant i\leqslant p-2$, such that $\beta=\alpha^i$. In
other words, $i=\log_{\alpha} \beta$. We set $\theta^{\log_{\alpha} 0}=0$ throughout
this paper.

Every sequence with period $p$ can be denoted by
$\varphi=(\varphi(0), \varphi(1), \cdots, \varphi(p-1))$, and also
considered as a vector in the Hilbert space
$\mathcal{H}=\mathbb{C}(\mathbb{F}_p)$ with the inner product given
by the standard formula: $<\varphi,\psi>=\sum_{i\in \mathbb{F}_p}
\varphi(i)\overline{\psi(i)}$ where $\overline{x}$ is the complex
conjugate of $x$. We denote $U(\mathcal{H})$ (Appendix 7.3) as the group of unitary
operators on $\mathcal{H}$. Let $L_t, M_w$ and $F$ be unitary
operators of the time-shift, phase-shift and DFT
respectively, which are defined as follows,

\begin{equation}\label{eq-un}
L_t[\varphi](i)=\varphi(i+t)\ \ \ \ \
M_w[\varphi](i)=\eta^{wi}\varphi(i)\ \ \ \ \mbox{and} \ \ \
F[\varphi](j)=\frac{1}{\sqrt{p}}\sum_{i\in
\mathbb{F}_p}\eta^{ji}\varphi(i), \,\, \varphi \in \mathcal{H}.
\end{equation}
We also use the notation  $\widehat{\varphi}$  for  $F[\varphi]$ for
simplicity. If   $\psi =L_t \varphi$ or $\psi=M_w \varphi$, then we
say that $\varphi$ and $\psi$ are {\em time-shift equivalent} or
{\em phase-shift equivalent}. Otherwise, they are {\em time-shift
distinct} or {\em phase-shift distinct}.

We denote $C_{\varphi}(t)$ and $C_{\varphi, \psi}(t)$ their
respective {\em autocorrelation} and {\em cross correlation}
functions, which are defined by
\begin{equation}
C_{\varphi}(t)=\sum_{i\in \mathbb{F}_p}
\varphi(i)\overline{\varphi(i+t)}\ \ \ \mbox{and}  \ \ \ C_{\varphi,
\psi}(t)=\sum_{i\in \mathbb{F}_p} \varphi(i)\overline{\psi(i+t)}.
\end{equation}

\begin{definition}
We say that $S$ is a $(p, r, \sigma,
\rho)$ {\em signal set} if each sequence in $S$ has period $p$,
there are $r$ time-shift distinct sequences in $S$, and the maximum
magnitude of out-of-phase autocorrelation  values and cross
correlation values are upper bounded by $\sigma$ and $\rho$
respectively, i.e.,

\begin{equation}\label{ac1}
\begin{array}{cl}
&|C_{\varphi}(t)| \leqslant \sigma,   \ \ t\neq 0, \varphi \in S, \\
&|C_{\varphi, \psi}(t)| \leqslant \rho,  t \in \F_p, \, \,  \varphi \neq \psi \in S.
\end{array}
\end{equation}

\end{definition}
In this paper,  we also say that  auto and cross correlation of $S$ is upper bounded
by $\sigma$ and $\rho$ respectively.

We say that a sequence $\varphi$ is a {\em perfect sequence}  if
\[
C_{\varphi}(t)=\left\{
\begin{array}{l l}
p &  t \equiv 0 \bmod{p},\\
0 &  t \notequiv 0 \bmod{p}.
\end{array}
\right.
\]

The {\em auto and cross ambiguity functions} of sequences are
defined as two-dimensional autocorrelation and cross correlation
functions in both time and phase, and are given by
\begin{equation}\label{eq-AF1}
A_{\varphi}(t, w) = <\varphi, M_wL_t\varphi>\,\, \mbox{and}\,\,\,
A_{\varphi, \psi}(t, w)=<\varphi, M_wL_t\psi>.
\end{equation}
The definitions of the auto and cross correlation
functions are equal to their respective auto and
cross ambiguity functions for the case $w=0$.

\begin{definition}
We say that $S$ is a $(p, r, \sigma,
\rho)$ {\em ambiguity signal set} if each sequence in $S$ has period $p$,
there are $r$ time-shift distinct sequences in $S$, and the maximum
magnitude of ambiguity out-of-phase autocorrelation values and cross
correlation values are upper bounded by $\sigma$ and $\rho$
respectively, i.e.,
\begin{equation}\label{ac1}
\begin{array}{cl}
&|A_{\varphi}(t, w)|\leqslant \sigma,\ \  (t,
w)\neq (0, 0),\\
&|A_{\varphi, \psi}(t, w)|\leqslant \rho,   \ \varphi \neq \psi \in S.
\end{array}
\end{equation}
 \end{definition}

\begin{property}
Let $S_1$ be a $(p, r, \sigma,
\rho)$ ambiguity signal set, and
$S_2=\{M_w \varphi| w\in \F_p, \varphi \in S_1\}$. Then $S_2$ is a $(p, pr, \sigma, \rho)$ signal set.
\end{property}

\begin{remark}
\em{ All the definitions and notations are stated for sequences
with period $p$ in this section. However, they are also valid for
sequences with period $n$ when $p$ and $\mathbb{F}_p$ are replaced
by $n$ and $\mathbb{Z}_n$ respectively.}
\end{remark}

\section{Main Results}

There are two types of sequences in the set of the
{\em finite oscillator system}  $\mathfrak{S}$ \cite{S.Gurevich}. One is from the split
case, denoted as $\mathfrak{S}^s$, and the other is from the non-split
case, denoted as $\mathfrak{S}^{ns}$. In other words,
$$\mathfrak{S}= \mathfrak{S}^s\cup  \mathfrak{S}^{ns}.$$
Gurevich, Hadani, and Sochen investigated how to implement
the sequences in $\mathfrak{S}^s$ by an algorithm \cite{S.Gurevich}.
Here we found a simple elementary construction for the
sequences in $\mathfrak{S}^s$, which is presented as follows.

\begin{theorem}
Let $\alpha$ be a generator of $\mathbb{F}_p^*$.
$$ \mathfrak{S}^s=\{\varphi_{x,y,z}\,|\, \ 1 \leqslant
x\leqslant p-2, 0\leqslant y\leqslant p-1, 0\leqslant z\leqslant
(p-1)/2\}$$ where $\varphi_{x,y,z}=\{\varphi_{x,y,z}(i)\}$ is a
normalized sequence with period $p$ whose elements are given by
$$\varphi_{x,y,0}(i)= \frac{1}{\sqrt{p-1}}\theta^{x\cdot \log_{\alpha}
i}\eta^{yi^2},$$ and
$$\varphi_{x,y,z}(i)=\frac{\eta^{yi^2}}{\sqrt{p(p-1)}}\sum_{j=1}^{p-1}\theta^{x\cdot
\log_{\alpha} j}\eta^{-(2z)^{-1}(j-i)^2} \ \mbox{for} \ z\neq 0.$$
\end{theorem}

If $z\neq 0$, it is clearly every element in $\varphi_{x,y,z}$ has
complicated form and does not lie on the unit circle, so we only consider the sequences where $z=0$.

\vspace{0.1in}

\noindent
{\bf Construction of  $\Omega_0$.}  Let $\alpha$ be a generator of $\mathbb{F}_p^*$.
For a given prime $p\ (p\geqslant 5)$, $n\in \mathbb{Z}$ and
$0\leqslant n<p(p-2)$, $n$ has a $p$-adic decomposition given by:
$n=(x-1)p+y$ where $1\leqslant x\leqslant p-2, 0\leqslant y
\leqslant p-1.$ Let $\varphi_n=\{\varphi_n(i)\}$ be a sequence  whose elements are defined as
$$\varphi_n(i)=\theta^{x\cdot \log_{\alpha} i}\cdot \eta^{yi^2},\ \ 0\leqslant i\leqslant p-1,$$
and $$\Omega_0=\{\varphi_n: 0\leqslant n<p(p-2)\}.$$
Then from the main results of \cite{S.Gurevich} (also Theorem 3 in this paper), we have

\begin{theorem}
The Signal set $\Omega_0$ satisfies the following properties.
\begin{enumerate}
\item[(a)] $\Omega_0$ is a $(p, p(p-2), 2\sqrt{p}, 4\sqrt{p})$ ambiguity signal set.
\item[(b)] DFT of $\varphi$ is bounded by $|\widehat{\varphi}(i)| < 2$, for $\varphi \in \Omega_0, i\in
\mathbb{F}_p$.
\item[(c)] The elements of each sequence $\varphi$ in $\Omega$ lie on the unit circle of the complex plane except $\varphi(0)=0$.
\end{enumerate}
\end{theorem}

We can extend  $\Omega_0$ by the phase shift operator as follows.

\vspace{0.1in}

\noindent
{\bf Construction of $\Omega$.} Let $\alpha$ be a generator of $\mathbb{F}_p^*$.
For a given prime $p\ (p\geqslant 5)$, $n\in \mathbb{Z}$ and
$0\leqslant n<p^2(p-2)$, $n$ has a $p$-adic decomposition given by:
$n=(x-1)p^2+yp+z$ where $1\leqslant x\leqslant p-2, 0\leqslant y,z
\leqslant p-1.$ Let $\varphi_n=\{\varphi_n(i)\}$ be a sequence  whose elements are defined as
$$\varphi_n(i)=\theta^{x\cdot \log_{\alpha} i}\cdot \eta^{yi^2+zi},\ \ 0\leqslant i\leqslant p-1,$$
and $$\Omega=\{\varphi_n: 0\leqslant n<p^2(p-2)\}.$$

Then from Property 1, we have

\begin{corollary}
The signal set $\Omega$ satisfies the following properties.
\begin{enumerate}
\item[(a)] $\Omega$ is a $(p, p^2(p-2), 2\sqrt{p}, 4\sqrt{p})$ signal set.
\item[(b)] DFT of $\varphi$ is bounded by $|\widehat{\varphi}(i)| < 2$, for $\varphi \in \Omega, i\in
\mathbb{F}_p$.
\item[(c)] The elements of each sequence $\varphi$ in $\Omega$ lie on the unit circle of the complex plane except $\varphi(0)=0$.
\item[(d)] The magnitude of auto ambiguity function of every sequence in $\Omega$ is upper bounded
by $2\sqrt{p}$ at any shift not equal to $(0, 0)$.
\end{enumerate}
\end{corollary}

\begin{example} {\em
For $p=5$, $a=2$ is a generator of $\mathbb{F}_5$,  the  elements of
the sequences $\varphi_x, \varphi_y,$ and  $\varphi_z$   are defined
as $\varphi_x(i)=\theta^{x\cdot\log_{\alpha} i}, \varphi_y(i)=\eta^{yi^2},$
and $\varphi_z(i)=\eta^{zi}$ respectively, which are given as
follows.

\begin{center}
\begin{tabular}{c c c}

\begin{tabular}{ |c| c |}
    \hline
     $x$  &   $\varphi_x(i)=\theta^{x\cdot\log_{\alpha} i}$        \\ \hline
     $1$  &   $\{0, 1, \theta,   \theta^3, \theta^2\}$ \\ \hline
     $2$  &   $\{0, 1, \theta^2,  \theta^2, 1  \} $    \\ \hline
     $3$  &   $\{0, 1, \theta^3,  \theta,   \theta^2\}$\\ \hline
\end{tabular}

&    \begin{tabular}{| c| c |}
    \hline
    $y$ &   $\varphi_y(i)=\eta^{yi^2}$       \\ \hline
    $0$ &   $\{1,\ 1,\ 1,\ 1,\ 1\ \}$             \\ \hline
    $1$ &   $\{1,\ \eta,\ \eta^4,\eta^4,\eta\}$       \\ \hline
    $2$ &   $\{1,\eta^2,\eta^3,\eta^3,\eta^2\}$     \\ \hline
    $3$ &   $\{1,\eta^3,\eta^2,\eta^2,\eta^3\}$     \\ \hline
    $4$ &   $\{1,\ \eta^4,\eta,\ \eta,\ \eta^4\}$     \\ \hline
   \end{tabular}

&    \begin{tabular}{| c| c |}
    \hline
        $z$ &   $\varphi_z(i)=\eta^{zi}$    \\ \hline
        $0$ &   $\{1,\ 1,\ 1,\ 1,\ 1\}$          \\ \hline
        $1$ &   $\{1,\eta,\eta^2,\eta^3,\eta^4\}$     \\ \hline
        $2$ &   $\{1,\eta^2,\eta^4,\eta,\eta^3\}$     \\ \hline
        $3$ &   $\{1,\eta^3,\eta,\eta^4,\eta^2\}$     \\ \hline
        $4$ &   $\{1,\eta^4,\eta^3,\eta^2,\eta\}$       \\ \hline
   \end{tabular}
\end{tabular}
\end{center}

Then the elements of each sequence in the signal set $\Omega$  are
constructed by   term-by-term products  of the elements of
$\varphi_x, \varphi_y,$ and $\varphi_z$. The first three sequences
and last two sequences are given as follows.

\begin{eqnarray*}
\varphi_0&=&\varphi_{1,0,0}=(0, 1, \theta, \theta^3,
\theta^2),\\
\varphi_1&=&\varphi_{1,0,1}=(0, \eta, \theta\eta^2, \theta^3\eta^3,
\theta^2\eta^4),\\
\varphi_2&=&\varphi_{1,0,2}=(0, \eta^2, \theta\eta^4, \theta^3\eta,
\theta^2\eta^3),\\
\vdots \ &&\ \ \ \ \ \ \ \ \ \ \ \ \vdots\\
\varphi_{73}&=&\varphi_{3,4,3}=(0, \eta^2, \theta^3\eta^2, \theta,
\theta^2\eta),\\
\varphi_{74}&=&\varphi_{3,4,4}=(0, \eta^3, \theta^3\eta^4,
\theta\eta^3, \theta^2).
\end{eqnarray*}
}
\end{example}

In the rest of the sections, we first prove  that Theorem 1 is the
split case of the finite oscillator system, and then
complete  proofs  for Theorem 2 and Corollary 1. In order to do so, in the next
section, we first  introduce some basic concepts and definitions on Weil representations,
and then present the oscillator system signal set.

\section{The Weil Representation and Finite Oscillator System}
For more details about the representation theory
and the Weil representation, we refer the reader to  \cite{S.Gurevich, S.D.Howard, R.Howe}
as well as the appendix in this paper.

\subsection{Weil Representation}

The  Weil representation is a unitary representation from $SL_2(\mathbb{F}_p)$
to $U(\mathcal{H})$ (see the  details in Appendix).
$SL_2(\mathbb{F}_p)$ can be generated by $g_a=\left(
\begin{array}{cc}
a & 0 \\
0 & a^{-1} \\
\end{array}
\right)$, $g_b=\left(
\begin{array}{cc}
 1 & 0 \\
 b & 1 \\
 \end{array}
 \right)$, and Weyl element
$w=\left(
\begin{array}{cc}
 0 & 1 \\
  -1 & 0 \\
  \end{array}
 \right)$
where $a\in \mathbb{F}_p^*$ and $b \in \mathbb{F}_p$. The Weil representations for $g_a,g_b$ and $w$ are
given in \cite{S.Gurevich2} as follows
\begin{equation}\label{eq-sc1}
\rho(g_a)[\varphi](i)=\sigma(a)\varphi(a^{-1}i)\\
\end{equation}
\begin{equation}\label{eq-ch1}
\rho(g_b)[\varphi](i)=\eta^{-2^{-1}bi^2}\varphi(i)
\end{equation}
\begin{equation}\label{eq-fo1}
\rho(w)[\varphi](j)=\frac{1}{\sqrt{p}}\sum_{i\in
\mathbb{F}_p}\eta^{ji}\varphi(i)
\end{equation}
where $\sigma: \mathbb{F}_p^*\rightarrow\{\pm 1\}$ is the Legendre
character, i.e., $\sigma (a)=a^{\frac{p-1}{2}}$ in $\mathbb{F}_p$.

Obviously, $\rho(w)$ is equal to  $F$
defined in (\ref{eq-un}). Here we denote $\rho(g_a)=S_a, \rho(g_b)=N_b,
\rho(w)=F$ for simplicity. For $g =\left(
\begin{array}{cc}
a & b \\
c & d \\
\end{array}
\right) \in SL_2(\mathbb{F}_p)$, if $b\neq 0$, we have
$$g=\left(
\begin{array}{cc}
 a & b \\
 c & d \\
 \end{array}
 \right)=
 \left(
\begin{array}{cc}
a & b \\
(ad-1)b^{-1} & d \\
\end{array}
\right)= \left(
\begin{array}{cc}
b & 0 \\
0 & b^{-1} \\
\end{array}
\right) \left(
\begin{array}{cc}
1 & 0 \\
bd & 1 \\
\end{array}
\right) \left(
\begin{array}{cc}
0 & 1 \\
-1 & 0 \\
\end{array}
\right) \left(
\begin{array}{cc}
1 & 0 \\
ab^{-1} & 1 \\
\end{array}
\right).$$ Thus the Weil representation of $g$ is given by
\begin{equation}
\rho(g)=S_{b}\circ N_{bd}\circ F\circ N_{ab^{-1}}.
\end{equation}
If $b=0$, then
$$g=\left(
\begin{array}{cc}
 a & b \\
 c & d \\
 \end{array}
 \right)=\left(
\begin{array}{cc}
 a & 0 \\
 c & a^{-1} \\
 \end{array}
 \right)=
 \left(
\begin{array}{cc}
 a & 0 \\
 0 & a^{-1} \\
 \end{array}
 \right)\left(
\begin{array}{cc}
 1 & 0 \\
 ac & 1 \\
 \end{array}
 \right).$$
Hence the Weil representation of $g$ is as follows
\begin{equation}
\rho(g)=S_a\circ N_{ac}.
\end{equation}

\subsection{The Finite Oscillator System}
In this subsection, we introduce the main results of
\cite{S.Gurevich}.

\vspace{0.1in} \noindent {\bf A. Maximal Algebraic Tori}

A {\em maximal algebraic torus} \cite{Borel} in $SL_2(\mathbb{F}_p)$ is a maximal commutative
subgroup which becomes diagonalizable over the original field or
quadratic extension of the field. One example of a maximal
algebraic torus in $SL_2(\mathbb{F}_p)$ is the standard diagonal
torus
$$A=\left\{\left(
\begin{array}{cc}
 a & 0 \\
 0 & a^{-1} \\
 \end{array}
 \right): a\in \mathbb{F}_p^*
\right\}.$$

Up to conjugation, there are two classes of the maximal algebraic
tori in $SL_2(\mathbb{F}_p)$. The first class, called {\em split
tori}, consists of those tori which are diagonalizable over
$\mathbb{F}_p$. Every split torus $T$ is conjugated to the standard
diagonal torus $A$, i.e., there exists an element $g \in
SL_2(\mathbb{F}_p)$ such that $g\cdot T \cdot g^{-1}= A$. The second
class, called {\em non-split tori}, consists of those tori which are
not diagonalizable over $\mathbb{F}_p$, but become diagonalizable
over the quadratic extension $\mathbb{F}_{p^2}$. In fact, a split
torus is a cyclic subgroup of $SL_2(\mathbb{F}_p)$ with order $p-1$,
while a non-split torus is a cyclic subgroup of $SL_2(\mathbb{F}_p)$
with order $p+1$.

All split (non-split) tori are conjugated to one another, so the
number of split (non-split) tori is the number of elements in the
coset space $SL_2(\mathbb{F}_p)/N$ ($SL_2(\mathbb{F}_p)/M$) (see
\cite{B. L. van der Waerden} for basics of group theory), where $N$
($M$) is the {\em normalizer group} of a non-split torus {\em A}.
Thus
\begin{equation}
\#(SL_2(\mathbb{F}_p)/N)=\frac{1}{2}p(p+1) \ \ \ \ \ \ \mbox{and}\ \
\ \ \ \ \ \#(SL_2(\mathbb{F}_p)/M)=\frac{1}{2}p(p-1).
\end{equation}
\begin{remark}
{\em A direct calculation shows that
the number of non-split tori is equal to  $\frac{1}{2}p(p-1)$ instead of  $p(p-1)$, which is a mistake made  in \cite{S.Gurevich}.}
\end{remark}

\vspace{0.1in} \noindent {\bf B. Decomposition of Weil
Representation Associated with Maximal Tori}

Because every maximal torus $T\in SL_2(\mathbb{F}_p)$ is a cyclic
group, restricting the Weil representation to $T$: $\rho_{|T}:
T\rightarrow U(\mathcal{H})$, we obtain a decomposition of $\rho_{|T}$  corresponding to an
orthogonal decomposition of $\mathcal{H}$.
\begin{equation}\label{eq-DE}
\rho_{|T}=\bigoplus_{\chi\in \Lambda_T}\chi\ \ \ \ \mbox{and} \ \ \
\mathcal{H}=\bigoplus_{\chi\in\Lambda_T}\mathcal{H}_\chi
\end{equation}
where $\Lambda_T$ is a collection of all the one dimensional
subrepresentations (characters) $\chi: T\rightarrow \mathbb{C}$ in the
decomposition of the Weil representation restricted on the torus $T$.

The decomposition (\ref{eq-DE}) depends on the type of $T$. In the
case where $T$  is a split torus, $\chi$ is a character given by
$\chi: \mathbb{Z}_{p-1}\rightarrow \mathbb{C}$. We have $dim
\mathcal{H}_{\chi} = 1$ unless $\chi=\sigma$ where $\sigma$ is the
Legendre character of $T$, and $dim \mathcal{H}_{\sigma} = 2$. In
the case where $T$ is a non-split torus, $\chi$ is the character
given by $\chi: \mathbb{Z}_{p+1}\rightarrow \mathbb{C}$. There is
only one  character $\sigma$ with order 2 that does not appear in the decomposition.
For the other characters $\chi\neq \sigma$,
$dim \mathcal{H}_{\chi} = 1$.

\vspace{0.1in} \noindent {\bf C. Sequences Associated with Finite
Oscillator System}

For a given torus $T$ and each character $\chi \in
\Lambda_T$, choosing a vector $\varphi_\chi\in
\mathcal{H}_\chi$ of unit norm, we obtain a collection of orthonormal vectors
\begin{equation}
\mathcal{B}_T=\{\varphi_\chi: \chi\in \Lambda_T, \chi\neq \sigma \}.
\end{equation}
Considering the union of these collections, then the finite oscillator system
\begin{equation}
\mathfrak{S}=\{\varphi\in \mathcal{B}_T: T\subset
SL_2(\mathbb{F}_p)\}.
\end{equation}

$\mathfrak{S}$ is naturally separated into two sub-systems
$\mathfrak{S}^{s}$ and $\mathfrak{S}^{ns}$ which correspond to the
split tori and the non-split tori respectively. The sub-system
$\mathfrak{S}^{s}$ ($\mathfrak{S}^{ns}$) consists of the union of
$B_T$, where $T$ runs through all the split tori (non-split tori) in
$SL_2(\mathbb{F}_p)$. Totally there are $\frac{1}{2}p(p+1)$ \
($\frac{1}{2}p(p-1)$) tori consisting of $p-2$ ($p$)
orthonormal sequences. Hence
\begin{equation}
\#\mathfrak{S}^{s}=\frac{1}{2}p(p+1)(p-2)\ \ \ \ \mbox{and} \ \ \
\#\mathfrak{S}^{ns}=\frac{1}{2}p^2(p-1).
\end{equation}

\begin{theorem}
Sequences in the set $\mathfrak{S}$ satisfy  the following
properties. For $ \varphi, \psi \in \mathfrak{S}$ and
$(t,w)\in V=\mathbb{F}_p\times \mathbb{F}_p$,
\begin{enumerate}
\item[(a)]  $\mathfrak{S}$ is a $(p, p(p^2-p-1),\frac{2\sqrt{p}}{p-1}, \frac{4\sqrt{p}}{p-1})$ ambiguity signal set.
\item[(b)] Supremum of $\varphi$ is given by  $ \max\{|\varphi(i)|:i\in
\mathbb{F}_p\}\leqslant \frac{2}{\sqrt{p}}$.
\item[(c)] For every sequences
$\varphi \in \mathfrak{S}$, its DFT $\hat{\varphi}$ is
(up to multiplication by a unitary scalar) also in $\mathfrak{S}$.
\end{enumerate}
\end{theorem}

\section{Proof of Main Results}
An efficient method to specify the decomposition (\ref{eq-DE}) is by
choosing a generator $t\in T$,  the character which is generated by the
eigenvalue of linear operator $\rho(t)$, and the character
space $\mathcal{H}_\chi$  that naturally corresponds to the eigenspace.  Below  are three steps to  construct  the sequences in the split case
of finite oscillator system $\mathfrak{S}^s$.

\begin{description}

\item[Step 1]  Compute the generator $g_{\alpha}$ for the standard torus $A$ and
$\mathcal{B}_A$. In other words, the collection of the eigenvectors
of $\rho(g_a)$ which do not correspond to eigenvalue $-1$.

\item[Step 2]  Compute all representative elements $g$ in the coset $\{
gN(A):g\in SL_2(\mathbb{F}_p)\}$ where $N(A)$ is the normalizer
group of $A$.

\item[Step 3] Compute all sequences $\rho(g)\varphi$ where $g$ is
the representative element presented in Step 2 and $\varphi\in
\mathcal{B}_A$ is calculated in Step 1.
\end{description}

Considering $\{\delta_i: i\in \mathbb{F}_p\}$ as the Kronecker delta function of Hilbert space
$\mathcal{H}=\mathbb{C}(\mathbb{F}_p)$ (i.e., $\delta_i$ is defined as $\delta_i(j)=\delta_{ij}$ for $\forall
j\in\mathbb{F}_p$), every
sequence $\varphi=\{\varphi(i)\}$ with period $p$ can be written as
$\varphi=\sum_{i=0}^{p-1}\varphi(i)\delta_i$.
Recall that $SL_2({\mathbb{F}_p})$ can be generated by  $g_a=\left(
\begin{array}{cc}
a & 0 \\
0 & a^{-1} \\
\end{array}
\right)$, $g_b=\left(
\begin{array}{cc}
 1 & 0 \\
 b & 1 \\
 \end{array}
 \right)$ and
$w=\left(
\begin{array}{cc}
 0 & 1 \\
  -1 & 0 \\
  \end{array}
 \right)$
where $a\in \mathbb{F}_p^*$ and $b \in \mathbb{F}_p$, then their
respective Weil representations (\ref{eq-sc1}), (\ref{eq-ch1}), and
(\ref{eq-fo1}) of $g_a$, $g_b$, and $w$ can be rewritten as follows
\begin{equation}\label{eq-sc2}
\rho(g_a)\delta_i=S_a\delta_i=\sigma(a)\delta_{ai}\\
\end{equation}
\begin{equation}\label{eq-ch2}
\rho(g_b)\delta_i=N_b\delta_i=\eta^{-2^{-1}bi^2}\delta_i
\end{equation}
\begin{equation}\label{eq-fo2}
\rho(w)\delta_j=F\delta_j=\frac{1}{\sqrt{p}}\sum_{i\in
\mathbb{F}_p}\eta^{ji}\delta_i.
\end{equation}

\begin{lemma}
Let $\alpha$ be a generator of $\mathbb{F}_p^*$, and $A=\left\{\left(
\begin{array}{cc}
 a & 0 \\
 0 & a^{-1} \\
 \end{array}
 \right): a\in \mathbb{F}_p^*
\right\}$ be the standard diagonal torus. Then
$$\mathcal{B}_A=\left\{\varphi_x=\frac{1}{\sqrt{p-1}}\sum_{i=1}^{p-1}\theta^{x\cdot
\log_{\alpha} i}\delta_i: 1\leqslant x \leqslant p-2 \right\}.$$
\end{lemma}
{\em Proof.} The set $\mathcal{B}_A$ is a collection of $\varphi_\chi$
with unit norm where $\varphi_\chi\in \mathcal{H}_\chi$  for every
character $\chi\ \neq\sigma$. In other words, the set
$\mathcal{B}_A$ is a collection of unit eigenvectors (not belonging to
eigenvalue $-1$) of $\rho(g_{\alpha})$  where $g_{\alpha}$ is a generator of Torus
$A$.

Let $\alpha$ be a generator of $\mathbb{F}_p^*$. Then $g_{\alpha}=\left(
  \begin{array}{cc}
    \alpha & 0 \\
    0 & \alpha^{-1} \\
  \end{array}
\right)$ is a generator of torus $A$. From (\ref{eq-sc2}), we have
$$\rho(g_{\alpha})\delta_i=\sigma(\alpha)\delta_{\alpha i}=-\delta_{ai}.$$
The eigenfunction of $\rho(g_{\alpha})$ is $(x+1)(x^{p-1}-1$), so the
eigenvalues of $\rho(g_{\alpha})$ are $-1, \theta^0,\theta^1, \theta^2
\cdots\cdots \theta^{p-2}$. Obviously, $-1=\theta^{\frac{p-1}{2}}$
occurs twice in the eigenvalues set. We assert that
$\sum_{i=1}^{p-1}\theta^{(\frac{p-1}{2}-j)\log_{\alpha} i}\delta_i$ is an
eigenvector associated to the eigenvalue $\theta^j (0\leqslant
j\leqslant p-2)$, and it can be verified as follows
\begin{eqnarray*}
\rho(g_{\alpha})(\sum_{i=1}^{p-1}\theta^{(\frac{p-1}{2}-j)\log_{\alpha}
i}\delta_i)&=&-\sum_{i=1}^{p-1}\theta^{(\frac{p-1}{2}-j)\log_{\alpha}
i}\delta_{ai}\\
&=&-\sum_{i=1}^{p-1}\theta^{(\frac{p-1}{2}-j)\log_{\alpha} (a^{-1} i)}\delta_i\\
&=&\theta^{\frac{p-1}{2}}\sum_{i=1}^{p-1}\theta^{(\frac{p-1}{2}-j)(\log_{\alpha} i-1)}\delta_i\\
&=&\theta^{\frac{p-1}{2}}\theta^{j-\frac{p-1}{2}}\sum_{i=1}^{p-1}\theta^{(\frac{p-1}{2}-j)\log_{\alpha}
i}\delta_i\\
&=&\theta^j\sum_{i=1}^{p-1}\theta^{(\frac{p-1}{2}-j)\log_{\alpha}
i}\delta_i.
\end{eqnarray*}
Let $x=\frac{p-1}{2}-j$. Then $\{\sum_{i=1}^{p-1}\theta^{x\cdot
\log_{\alpha} i}\delta_i\ (1\leq x\leq q-2)\}$ is a set of the
eigenvectors corresponding to all the eigenvalues not equal to $-1$.
By normalizing the eigenvectors, we complete the proof. \done

\begin{lemma}
Let $A=\left\{\left(
\begin{array}{cc}
 a & 0 \\
 0 & a^{-1} \\
 \end{array}
 \right): a\in \mathbb{F}_p^*
\right\}$ be the standard diagonal torus, and $N(A)$ be the normalizer group of $A$. Then
$$R=\left\{\left(
\begin{array}{cc}
1 & b \\
c & 1+bc \\
\end{array}
\right): 0 \leqslant b \leqslant \frac{p-1}{2}, c\in
\mathbb{F}_p\right\}$$ is a collection of coset representatives of $\{ gN(A):g\in
SL_2(\mathbb{F}_p)\}$.
\end{lemma}
{\em Proof.} Denote $B=\left\{\left(
\begin{array}{cc}
 0 & -b \\
 b^{-1} & 0 \\
 \end{array}
 \right): b\in \mathbb{F}_p^*
\right\}$. Then it's not hard to verify
$$N(A)=\{g: gAg^{-1}=A, g\in
SL_2(\mathbb{F}_p)\}=AB.$$ Thus every representative element $g$ can
be written as the form $$g=\left(
\begin{array}{cc}
1 & b \\
c & 1+bc \\
\end{array}
\right)\ \ b,c\in \mathbb{F}_p.$$
Note that $g=\left(
\begin{array}{cc}
1 & b \\
c & 1+bc \\
\end{array}
\right)$ and $g'=\left(
\begin{array}{cc}
1 & b' \\
c' & 1+b'c'\\
\end{array}
\right)$ are in the same coset, i.e., $g^{-1}g'\in N(A)$, if and only if
\begin{eqnarray*}
\left(
\begin{array}{cc}
1 & b' \\
c' & 1+b'c'\\
\end{array}
\right)&=&\left(
\begin{array}{cc}
1 & b\\
c & 1+bc \\
\end{array}
\right) \left(
\begin{array}{cc}
 0 & -b \\
 b^{-1} & 0 \\
 \end{array}
\right)\\
&=&
 \left(
   \begin{array}{cc}
     1 & -b \\
     b^{-1}+c & -bc \\
   \end{array}
 \right)\\
 &=&
 \left(
   \begin{array}{cc}
     1 & -b \\
     b^{-1}+c & 1+(-b)(b^{-1}+c) \\
   \end{array}
 \right)
\end{eqnarray*}
if and only if $b'=-b$ and  $c'=b^{-1}+c$. Therefore $R$ contains all
representative elements in the coset $\{ gN(A):g\in
SL_2(\mathbb{F}_p)\}$.

 \done

By Lemmas 1 and 2, we can now prove Theorem 1, which is a direct consequence of the following result.

\begin{proposition} There are two types of vectors in $\mathfrak{S}^s$. \\
The first type is
$$\varphi_{x,y,0}=\frac{1}{\sqrt{p-1}}\sum_{i=1}^{p-1}\theta^{x\cdot \log_{\alpha} i}\eta^{yi^2}\delta_i$$
where $1 \leqslant x\leqslant p-2, 0\leqslant y\leqslant p-1.$\\
The second type is
$$\varphi_{x,y,z}=\frac{1}{\sqrt{p(p-1)}}\sum_{i=0}^{p-1}\sum_{j=1}^{p-1}\theta^{x\cdot \log_{\alpha}
j}\eta^{yi^2-(2z)^{-1}(j-i)^2}\delta_i$$ where $1 \leqslant
x\leqslant p-2, 0\leqslant y\leqslant p-1, 1\leqslant z\leqslant
\frac{p-1}{2}.$
\end{proposition}

{\em Proof.} Every split torus $T\subset SL_2({\mathbb{F}_p})$ can be
written as the form $gAg^{-1}$ where $A$ is the diagonal torus and
$g=\left(
                                                    \begin{array}{cc}
                                                      1 & b \\
                                                      c & 1+bc \\
                                                    \end{array}
                                                  \right)
\in R$ in Lemma 2. Then
$$\mathcal{B}_T=\mathcal{B}_{gAg^{-1}}=\{\rho(g)\varphi: \varphi\in
\mathcal{B}_A\},$$ and
$$\mathfrak{S}^s=\bigcup_{g\in R} \mathcal{B}_{gTg^{-1}}=\{\rho(g)\varphi: g\in R, \varphi\in
\mathcal{B}_A\}.$$ If $b=0$, $g=\left(
  \begin{array}{cc}
    1 & b \\
    c & 1+bc \\
  \end{array}
\right)$ has the form $\left(
\begin{array}{cc}
1 & 0 \\
c & 1 \\
\end{array}
\right) (0\leqslant c\leqslant p-1)$, then from (\ref{eq-ch2}), we
have
\begin{eqnarray*}
\rho(g)\varphi_x&=&N_c(\frac{1}{\sqrt{p-1}}\sum_{i=1}^{p-1}\theta^{x\cdot
\log_{\alpha} i}\delta_i)\\
&=&\frac{1}{\sqrt{p-1}}\sum_{i=1}^{p-1}\theta^{x\cdot \log_{\alpha}
i}N_c\delta_i\\
&=&\frac{1}{\sqrt{p-1}}\sum_{i=1}^{p-1}\theta^{x\cdot \log_{\alpha}
i}\eta^{-2^{-1}ci^2}\delta_i.
\end{eqnarray*}
If $b\neq 0$, $g$ has the following decomposition
$$g=\left(
  \begin{array}{cc}
    1 & b \\
    c & 1+bc \\
  \end{array}
\right)= \left(
\begin{array}{cc}
b & 0 \\
0 & b^{-1} \\
\end{array}
\right) \left(
\begin{array}{cc}
1 & 0 \\
b(1+bc) & 1 \\
\end{array}
\right) \left(
\begin{array}{cc}
0 & 1 \\
-1 & 0 \\
\end{array}
\right) \left(
\begin{array}{cc}
1 & 0 \\
b^{-1} & 1 \\
\end{array}
\right).$$ Then applying (\ref{eq-sc2}),(\ref{eq-ch2}), and
(\ref{eq-fo2}), for $1\leqslant x\leqslant p-1$, we have
\begin{eqnarray*}
\rho(g)\varphi_x&=&S_b\circ N_{b(1+bc)}\circ F\circ
N_{b^{-1}}(\frac{1}{\sqrt{p-1}}\sum_{j=1}^{p-1}\theta^{x\cdot \log_{\alpha}
j}\delta_j)\\
&=&S_b\circ N_{b(1+bc)}\circ
F(\frac{1}{\sqrt{p-1}}\sum_{j=1}^{p-1}\theta^{x\cdot \log_{\alpha}
j}\eta^{-2^{-1}b^{-1}j^2}\delta_j)\\
&=&S_b\circ
N_{b(1+bc)}(\frac{1}{\sqrt{p(p-1)}}\sum_{i=0}^{p-1}\sum_{j=1}^{p-1}\theta^{x\cdot
\log_{\alpha} j}\eta^{-2^{-1}b^{-1}j^2}\eta^{ij}\delta_i)\\
&=&S_b(\frac{1}{\sqrt{p(p-1)}}\sum_{i=0}^{p-1}\sum_{j=1}^{p-1}\theta^{x\cdot
\log_{\alpha} j}\eta^{-2^{-1}b^{-1}j^2}\eta^{ij}\eta^{-2^{-1}b(1+bc)i^2}\delta_i)\\
&=&\sigma(b)(\frac{1}{\sqrt{p(p-1)}}\sum_{i=0}^{p-1}\sum_{j=1}^{p-1}\theta^{x\cdot
\log_{\alpha} j}\eta^{-2^{-1}b^{-1}j^2}\eta^{ij}\eta^{-2^{-1}b(1+bc)i^2}\delta_{bi})\\
&=&\sigma(b)(\frac{1}{\sqrt{p(p-1)}}\sum_{i=0}^{p-1}\sum_{j=1}^{p-1}\theta^{x\cdot
\log_{\alpha}
j}\eta^{-2^{-1}b^{-1}j^2}\eta^{b^{-1}ij}\eta^{-2^{-1}b^{-1}(1+bc)i^2}\delta_i)\\
&=&\frac{\sigma(b)}{\sqrt{p(p-1)}}\sum_{i=0}^{p-1}\sum_{j=1}^{p-1}\theta^{x\cdot
\log_{\alpha} j}\eta^{-(2b)^{-1}(j-i)^2-2^{-1}ci^2}\delta_i.
\end{eqnarray*}
Let $y=-2^{-1}c, z=b$. Then $y$ ranges over $\mathbb{F}_p$ as $c$ ranges over $\mathbb{F}_p$. Note that $\sigma(z)= \pm 1$ is a
constant, so $\frac{1}{\sqrt{p-1}}\sum_{i=1}^{p-1}\theta^{x\cdot
\log_{\alpha} i}\eta^{yi^2}\delta_i$ and
$\frac{1}{\sqrt{p(p-1)}}\sum_{i=0}^{p-1}\eta^{yi^2}\sum_{j=1}^{p-1}\theta^{x\cdot
 \log_{\alpha} j}\eta^{-(2z)^{-1}(j-i)^2}\delta_i$ with $1 \leqslant
x\leqslant p-2$, $0\leqslant y\leqslant p-1$, $1\leqslant z\leqslant
\frac{p-1}{2}$ are all the vectors in $\mathfrak{S}^s$, which
completes the proof. \done

Thus, we have found a simple elementary representation for
the split case of the finite oscillator system.

The following lemma gives the relationship of the correlation function, ambiguity function, and unitary operator $L_t, M_w, F$ defined in
(\ref{eq-un}), which  is easy to verify.
\begin{lemma}
$\forall \varphi, \psi$ sequences with period $p$,
$\forall t, w, z \in \mathbb{F}_p$, and where $L_t, M_w, F$ are defined in
(\ref{eq-un}), we have:
\begin{enumerate}
\item[(a)] $ C_{\varphi}(t)=<\varphi, L_t\varphi>\ and \  C_{\varphi,
\psi}(t)=<\varphi, L_t\psi>.$
\item[(b)] $|<\varphi, \pi(t,w,z)\psi>|=|<\varphi, M_w \cdot
L_t\psi>|=|<\varphi, L_t \cdot M_w\psi>|.$
\item[(c)] $L_t\cdot F=F\cdot M_t$\ \ \mbox{and}\ \  $FL_{-t}=M_t\cdot F.$
\item[(d)]$<\widehat{\varphi}, L_t\widehat{\psi}>=<\varphi, M_{-t}\psi>
\ \ \mbox{and} \ \ <\widehat{\varphi}, M_w\widehat{\psi}>=<\varphi,
L_t\psi> (Parseval \ \ Formulae).$
\end{enumerate}
\end {lemma}

Now we extend signal set from $\mathfrak{S}$ to $\overline{\mathfrak{S}}$ by the phase shift operator, i.e.,
$$\overline{\mathfrak{S}}=\{M_w\varphi:\ \  \forall\varphi\in \mathfrak{S}, w\in
\mathbb{F}_p\}.$$ Then  $\overline{\mathfrak{S}}$ satisfy the following property.

\begin{property} With the above notation,
\begin{enumerate}
\item[(a)] $\overline{\mathfrak{S}}$ is a $(p, p^2(p^2-p-1),\frac{2\sqrt{p}}{p-1}, \frac{4\sqrt{p}}{p-1})$ signal set.
\item[(b)] Supremum of $\psi$ is given by  $ \max\{|\psi(i)|:i\in
\mathbb{F}_p\}\leqslant \frac{2}{\sqrt{p}}$, $\psi\in \overline{\mathfrak{S}}$.
\item[(c)] DFT of $\psi$ is bounded by $|\widehat{\psi}(i)|\leqslant \frac{2}{\sqrt{p}}, \forall  i\in
\mathbb{F}_p$,  $\psi\in \overline{\mathfrak{S}}$.
\end{enumerate}
\end {property}

{\em Proof.}
\begin{enumerate}
\item[(a)]  By Property 1, $\mathfrak{S}$ is a $(p, p(p^2-p-1),\frac{2\sqrt{p}}{p-1}, \frac{4\sqrt{p}}{p-1})$ ambiguity signal set, so
$\overline{\mathfrak{S}}$ is a $(p, p^2(p^2-p-1),\frac{2\sqrt{p}}{p-1}, \frac{4\sqrt{p}}{p-1})$ signal set.

\item[(b)]  $\forall \ M_w\varphi \in \overline{\mathfrak{S}}$, it is clear
that the magnitude of $M_w\varphi(i)$ is as same as that of $\varphi(i)$.

\item[(c)]  Applying Lemma 3-(c), the DFT of $M_w\varphi$ can be
written as $F\cdot M_w \varphi=L_w\cdot F\varphi$.  We can see  that
$F\varphi$ is also in $\mathfrak{S}$ from Lemma 3-(c), and $|F\varphi(i)|
\leqslant \frac{2}{\sqrt{p}}$ from Theorem 3-(b).  Thus $|F\cdot M_w
\varphi(i)|=|L_w\cdot F\varphi(i)|\leqslant \frac{2}{\sqrt{p}}$,
which completes the proof.
\end{enumerate}
\done

\vspace{0.1in} \noindent {\em Proof of  Theorem 2.} Considering $\Omega_0$ and  $\mathfrak{S}^s$, it is obvious that
$\Omega_0$ is a subset of $\mathfrak{S}^s$ up to
multiplication by $\sqrt{p-1}$. Thus $\Omega_0$ is a $(p, p^2(p-2),
2\sqrt{p}, 4\sqrt{p})$ ambiguity signal set, and the DFT of $\varphi\in \Omega_0$ is bounded by $|\widehat{\varphi}(i)|
\leqslant 2\sqrt{\frac{p-1}{p}}<2$  $\forall \varphi \in \Omega$ and $\forall  i\in
\mathbb{F}_p$. \done

From Theorem 2, Property 1 and Lemma 3, Corollary 1 holds.

\section{Comparisons of the New Constructions with Some Known Constructions}

The split case of the finite oscillator
system $\mathfrak{S}^s$ and the extended construction
$\overline{\mathfrak{S}}$ can be efficiently
implemented for  moderate $p$. However, for large $p$, since one
needs  to compute the exponential sum of $p$ elements, they are not so
efficient. Therefore, in this section, we only make some comparisons
for  the set $\Omega$ or $\Omega_0$  with some known constructions.

\vspace{0.1in} \noindent {\bf A. Compared with Complex Valued Sequences with
 Good Ambiguity Function or  DFT}

Let $n$ be a positive integer and  $\omega_n$ be an $n$th primitive root of unit in the complex field, i.e., $\omega_n=e^{-\frac{2\pi j}{n}}$ where $j=\sqrt{-1}$.
For fixed $0<y<n, 0\leqslant z<n$ where $y$ is relatively prime to $n$, a Frank-Zadoff-Chu (FZC)
sequence \cite{FrankZadoff62, Chu72, Frank73}  $\{\varphi_{y, z}(i)\}$ is  given by
\begin{equation}\label{eq-FZC}
\varphi_{y, z}(i)=\left\{
\begin{array}{ll}
\omega_n^{(1/2)y i^2+zi} & \mbox{$n$ is even},\\
&\\
\omega_n^{(1/2) y i(i+1)+zi} & \mbox{$n$ is odd}.\\
\end{array}
\right.
\end{equation}
Any FZC sequence is a perfect sequence, i.e., its  out-of-phase autocorrelation is zero. Note that $\omega_n^{1/2}$ is a ($2n$)th primitive root of unit in the complex field. For $n$ odd,  $\varphi_{y, z}(i)$ can be given by an equivalent expression: $\omega_n^{y'i^2+z'i}$ where $0<y'<n, 0\leqslant z'<n$.  This form will be used below.

\begin{enumerate}

\item For a fixed $z$, a FZC signal set is  a set consisting of the $\varphi(n)$ sequences defined by (\ref{eq-FZC}) where $\varphi(n)$ is the Euler function.   When $n=p$ a prime,  the FZC set is a $(p, p-1, 0, \sqrt{p})$ signal set. The magnitude
of the DFT of these sequences is bounded by 1.

\item The elements in Alltop cubic sequences \cite{Alltop80} with period
$p$ are given by $\varphi_y(i)=\omega_p^{i^3+yi}$ where $0\leqslant y
\leqslant p-1$. The auto and cross ambiguity function can reach $p$
with $\frac{1}{p}$ probability, and the magnitude of the DFT of these sequences is  bounded by $2$.

\item Sequences from Heisenberg representation:
The elements in a sequence from the Heisenberg representation
\cite{S.D.Howard} have the form $\varphi_{y,z}(i)=\omega_p^{yi^2 +zi}$
where $0\leqslant y,z \leqslant p-1$. (Note that the sequences from Heisenberg representation are the same as the FZC sequences with period  $n=p$, a prime.) Here the magnitude of the auto
ambiguity function of such sequences can reach $p$ with
$\frac{1}{p}$ probability, while the upper bound
of the cross ambiguity function between two phase-shift distinct
sequences is given by $\sqrt{p}$, and the magnitude of the DFT of these sequences is bounded by $1$.

\item Modulatable  orthogonal sequences \cite{Suehiro88}:
An $h \times h$ discrete Fourier transform (DFT) matrix is defined by the $j$th row and the $k$th column elements
of
\begin{equation}\label{eq:dft_exp}
d_{z, j, k} = \omega_h^{zjk}
\end{equation}
where $z$ is a fixed number with $0<z<h$ and $\gcd(z, h)=1$, and $0 \leqslant j, k < h$. Let a sequence $\{a_z(i)\}$ be given by  concatenation of the rows of DFT matrix starting from the first row, second row, and so on, i.e., $a_z(i=jh+k)=d_{z, j, k}, 0\leqslant j, k<h$. (Note that $\{a_z(i)\}$ can be considered as an interleaved sequence associated with the DFT matrix \cite{GolombGong-book05}.)
Let  $\{b(i)\}_{i\ge 0}$ be a complex valued sequence with period $h$ and $|b_i|=1$, i.e., the magnitude of $b_i$ is equal to 1.  A modulatable  orthogonal (MO)  sequence $\{c_z(i)\}$ of period $n=h^2$  is given by
\[
c_z(i)=a_z(i)b(i), i=0 ,1, \cdots.
\]
For each $h$, an MO sequence is a perfect sequence.  An  MO signal set consists of the sequences for all $z$.  When $h=p$, a prime, this set  is a signal set with  parameters $(p^2, p-1, 0,  p)$.
\item
Generalized chirp-like (GCL) sequences  \cite{Pop92}:  Popovi\'c,  generalized the construction of the modulatable orthogonal sequences in 1992 as follows. Let $\{\varphi_{y, z} (i) \}$ be a FZC sequence with period $n=th^2$ where both $t$ and $h$ are arbitrary positive integers, and  $\{b(i)\}$ be the same as defined for MO sequences.    A   generalized chirp-like sequence $\{c_{y, z}(i)\}$ is given by
\[
c_{y, z}(i)=\varphi_{y, z}(i)b(i),  i=0, 1, \cdots
\]
where the index $i$ of $\varphi_{y, z}(i)$ is reduced  modular $n$ and the index of $b(i)$ is reduced  modular $h$.  Each generalized chirp-like sequence sequence  is a perfect sequence.  For a fixed $z$,   a GCL signal set consists of all $\{c_{y, z}(i)\}$  for  GCD$(y, n)=1$. When $n=p^2$ where $p$ is a prime,  a GCL signal set is  a $(p^2, p-1, 0, p)$ signal set.

Note that their respective auto/cross ambiguity functions and the DFT of  MO and GCL sequences are not reported in the literature.   A more recent work \cite{Brodzik09} using the Zak transform  showed that the above perfect sequences, i.e., FZC, MO and GCL sequences,  can be considered as subsets of the sequences constructed from the Zak transform for some special parameters.

\item Power residue sequences \cite{Sidel69, Lempel77, Sarwate78}:  Let $k$ be a proper factor of $p-1$.  A power residue sequence $\{\varphi_x(i)\}$ of period $p$ is defined as
\begin{equation}\label{eq-P}
\varphi_{x}(i)=\omega_k^{x\cdot \log_a i}, i=0, 1, \cdots,
\end{equation}
where $0<x<k$.   A power residue sequence is a polyphase sequence with period $p$ and $k$ different phases, which is represented by multiplicative  characters.  A $k$-ary \emph{power residue sequence} of period $p$ has the out-of-phase autocorrelation magnitude of at most $3$, which  is also studied in \cite{Green}.  Moreover, it is shown in~\cite{Kim06} that the magnitude of the cross-correlation of distinct $k$-ary power residue sequences of period $p$ is bounded by $\sqrt{p}+2$.  Thus, the set consisting of the power residue sequences defined by (\ref{eq-P}) for all $x: 0<x<k$ is a signal set with parameters $(p, k-1, 3, \sqrt{p}+2)$ where $k$ can be up to $k=p-1$. When $k=p-1$, it  can be seen that this is a subset of  $\Omega$, the new expression of the sequences from the Weil  representation. Thus the ambiguity and the DFT are bounded with the same values as for $\Omega$.  Furthermore,  this signal set can be enlarged using the shift-and-add operators. For details, see a recent paper \cite{YuGong10}.

\item  For the new construction $\Omega$, there are $p^2(p-2)$ time-shift
distinct sequences, and the elements in every sequence have the expression
$\varphi_{x,y,z}(i)=\omega_{p-1}^{x\cdot \log_a i}\cdot \omega_p^{yi^2+zi}$ (note $\theta=\omega_{p-1}$ and $\eta=\omega_p$ in the previous notation for the new construction.
The magnitude of auto and cross correlaton of sequences in the set are upper bounded by $2\sqrt{p}$ and
$4\sqrt{p}$, respectively, and the magnitude of the DFT of these sequences is  upper bounded by $2$.   The subset $\Omega_0$ where $z=0$ is an ambiguity signal set with parameters $(p, p(p-2), 2\sqrt{p}, 4\sqrt{p})$. However, there is a possible drawback of those sequences in practice.
The alphabet for a sequence of length $p$ grows roughly as
$O(p^2)$.
\end{enumerate}

We summarize the above discussions in Table \ref{tab1}.  We use the notation $\eta=\omega_p$ as we used in the previous sections except for the case of GCL where we use  $\omega_{p^2}$.
\begin{table}[h!]
  \begin{center}
  \caption{The Comparison with Well-known Complex Valued Sequences}
\label{tab1}
    \begin{tabular}{| l|| c| c |c |l| }
    \hline
    Family & $i$th element   & Period $L$ & Size   & Ambiguity and DFT  \\ \hline
    \hline
    FZC$^{(1)}$                 &$\varphi_{y}(i)=\eta^{yi^2}$          &  &   & $|AA|: p$.     \\
           \cite{FrankZadoff62} \cite{Chu72} \cite{Frank73}       & $(0\leqslant y\leqslant p-1)$ & $p$    &$p$&  $|CA|\leqslant \sqrt{p}$. \\
                    &                 &   &     & $|DFT|\leqslant 1$.\\
                \hline
                                     &$\varphi_y(i)=\eta^{i^3+yi^2}$    &  &     & $|AA|: p$.        \\
    Alltop cubic  \cite{Alltop80}    &$ (0\leqslant y\leqslant p-1)$   & $p$    &$p$  &  $|CA|: p$.\\
    &&&& $|DFT|\leqslant 2$.
                                      \\ \hline
    Sequences from                   &$\varphi_{y}(i)=\eta^{yi^2+zi}$    &    &     & $|AA|: p$.  \\

    Heisenberg                       &$ (0\leqslant y \leqslant p-1)$  & $p$   &$p $&   $|CA|\leqslant \sqrt{p}$. \\
    representation \cite{S.D.Howard} &                                    &  &   &$|DFT|\leqslant 1$.   \\ \hline

    MO$^{(1)}$ \cite{Suehiro88} & $c_z(ip+k)=\eta^{zik} b(k)$ & $p^2$ & $p-1$ &AA, CA, DFT\\
   &                        ($1\leqslant z\leqslant p-1 $) &&& are unknown.\\ \hline

    GCL$^{(1)}$ \cite{Pop92} & $c_y(i)=\omega_{p^2}^{yi^2+zi} b(i)$  & $p^2$ & $p-1$ &AA, CA, DFT\\
    &                             ($1\leqslant y \leqslant p-1$ ) &&& are unknown.\\ \hline
    Power residue sequences & $\varphi_{x}(i)=\theta^{x\cdot \log_a i}$ & $p$ & $p-2$&  The same as $\Omega_0$.  \\
    \cite{Sidel69}\cite{Kim06} & $(1\leqslant x \leqslant p-2)$ &&& \\ \hline
        Sequences  from                &$\varphi_{x,y,z}(i)=\theta^{x\cdot \log_a i}\eta^{yi^2}$  & & & $|AA|\leqslant 2\sqrt{p}.$  \\
     Weil   representation                      &$ (1\leqslant x \leqslant p-2,  $ & $p$   &$p(p-2)$& $|CA|\leqslant 4\sqrt{p}.$      \\
     $\Omega_0$ (this paper)& $0\leqslant y \leqslant p-1)$ &&& $|DFT|\leqslant 2$ \\ \hline
  \end{tabular}
\end{center}
{\footnotesize
\begin{itemize}
\item[-] AA =Auto ambiguity, CA = Cross ambiguity.
\vspace{-0.05in}
\item[-]  $^{(1)}$ Those are perfect sequences.
\end{itemize}
}
\end{table}

 \vspace{0.1in} \noindent {\bf B. Signal Sets with  Sizes in the Order of $p^3$ and Low Correlation}

Signal sets with  family size in the  order of  $p^3$,  and
with low correlation are known in the literature  and are shown in Table \ref{tab2}. The
bounds of  auto and cross correlation function for construction  $\Omega$ are better than or as good as the sequences in
\cite{BlakeMark82}, $\Z_4$ sequences $S(2)$ \cite{Z4awarded}, and
the sequences in \cite{YuGong06}, while the maximum magnitudes of DFT are only known for $\Omega$, and $\Z_4$ sequences $S(2)$.

\begin{table}[h!]
  \begin{center}
  \caption{The Comparison with Sequences with Low Correlation}
  \label{tab2}
    \begin{tabular}{| c|| c |c |c| c|c|}
    \hline
    Family & Period $L$ &Size  & Correlation       & DFT  & Ambiguity      \\
    \hline \hline
    Blake and Mark \cite{BlakeMark82}$^{(2)}$ &   $p-1$     & $(L+1)^3$       &  $4\sqrt{L+1}+1$             & N   & N   \\ \hline
    $\Z_4$ sequences $S(2)$ \cite{Z4awarded}  &   $2^k-1$   & $L^3+4L^2+5L+2$ &  $4\sqrt{L+1}+1$             &5 \cite{PatersonTarokh00}  &  N  \\ \hline
    Yu and Gong           \cite{YuGong06}    &   $2^k-1$   & $(L+1)^3$       &  $2^{2.5}\sqrt{L+1}$           & N   & N \\ \hline
    $\Omega$                                 &   $p$       & $L^2(L-2)$      &  $2\sqrt{L}, 4\sqrt{L}$  & $2$  &  Not good \\ \hline
    $\Omega_0$                                 &   $p$       & $L(L-2)$      &  $2\sqrt{L}, 4\sqrt{L}$  & $2$  &  $|AA|\leqslant 2\sqrt{L}$, $|CA|\leqslant 4\sqrt{L}$  \\ \hline
   \end{tabular}
   \end{center}

  {\footnotesize
\begin{itemize}
\item[-] {$^{(2)}$ This family can be easily extended to sequences over the finite field $\F_p$ with period
  $p^n-1$ and the same correlation property from the work in
  \cite{Moreno94}.}
  \vspace{-0.05in}
  \item[-] AA =Auto ambiguity, CA = Cross ambiguity.
 \vspace{-0.05in}
\item[-] N: no reported results in the literature.
\end{itemize}
}

  \end{table}

 \vspace{0.1in} \noindent {\bf C. Implementation Cost}

Note that the $i$th  element  of a sequence  in $\Omega$ is a product of the $i$th element  of a $(p-1)$-ary power residue sequence of period $p$ and the $i$th element of an FZC sequence of period $p$.   Thus, the    implementation cost of construction $\Omega$  is equal to the sum of the cost of  those two types of  sequences.   Since both  power residue sequences and FZC sequences can be implemented efficiently at both hardware and software level,  so do  the sequences in  $\Omega$.    Furthermore,  the new expression of Weil representation sequences provides a trade-off among the alphabet size and good ambiguity.

\section{Concluding Remarks and An Open Problem}

We have discovered a simple elementary representation of the
sequences in the finite oscillator system from the Weil
representation, introduced by Gurevich, Hadani, and Sochen. From
this, we have shown a construction $\Omega$ of families of complex valued sequences of period $p$ having low
valued correlation functions. This
construction produces a signal set with  $p^2(p-2)$ shift
distinct sequences. The magnitude of the auto and cross correlation
functions are upper bounded by $2\sqrt{p}$ and $4\sqrt{p}$, respectively.  The DFT of every sequence in the signal set is upper bounded by 2. The signal set  $\Omega_0$, a subset of $\Omega$,  possesses all the properties of $\Omega$ as well as the magnitude of  their  auto and cross ambiguity functions are bounded by   $2\sqrt{p}$ and $4\sqrt{p}$.
However, there is a drawback of this construction in practice, since
the alphabet for a sequence of length p grows roughly as
$O(p^2)$.

If we look at the construction $\Omega$ again,  we find that each sequence
$\varphi_{n}=\{\varphi_{n}(i)\}_{i\geqslant 0}$  is the
term-by-term product of sequences $\{\theta^{x \log_{\alpha} i}
\}_{i\geqslant 0}$ and  $\{\eta^{yi^2 +zi}\}_{i\geqslant 0}$ which are related to power residue sequences and FZC sequences,
respectively.  Going back to the literature, all the known
constructions only involve one type of character from finite
field $\F_p$, While here we use both multiplicative and additive
characters of finite field $\F_p$.  The proof of
those results requires very deep mathematics, i.e., the
representation theory and $l$-adic algebraic geometry.
This suggests that it is worth looking for a direct proof
for the construction, which will have a two-fold
effect. One is for better promotion of those sequences in practice
without introducing the Weil representation theory. The other is
that it may lead to more discoveries of new signal sets with good
auto and cross ambiguity functions as well as with low magnitude of the
DFT spectrum.

\vspace{0.1in} \noindent {\bf Open Problem.} For $\Omega=\{\varphi_n
\,| \, 0\leqslant n \leqslant p^2(p-2)\}$, directly show that
$\Omega$ is a $(p,  p^2(p-2), 2\sqrt{p}, 4\sqrt{p})$ signal set and that the DFT of  every sequence is
upper bounded by 2 without introducing Weil representation and  finite
oscillator system.

\section*{Acknowledgment}
The authors would like to thank Grevich, Hadani and Sochen for their
help  during the course of conducting this work.  The authors
are deeply grateful to the Associate Editor and reviewers for their  many valuable and helpful suggestions which greatly improved the presentation of the work.

\section*{Appendix}
\subsection*{The Heisenberg Representation}

Let $V=\mathbb{F}_p^2$ be a two-dimensional vector space over
the finite field $\mathbb{F}_p$. Then $(V ,\omega)$ is symplectic if the symplectic form
$\omega$ is given by
$$\omega((t_1,w_1),(t_2,w_2))=t_1w_2-t_2w_1,$$
for $  (t_i,w_i)\in
V$, $i=1,2.$

Considering $V$ as an Abelian group, it admits a non-trivial central
extension called the {\em Heisenberg group} $H$ ($p\neq 2$). The
group $H$ can be presented as $H = V\times F_p$ with the
multiplication given by
$$(t_1,w_1,z_1)\cdot(t_2,w_2,z_2)=(t_1+t_2,w_1+w_2,z_1+z_2+2^{-1}\omega((t_1,w_1),(t_2,w_2))).$$
It is easy to verify that the center of $H$ is $Z = Z(H) = \{(0,0, z) :
z\in \mathbb{F}_p$\}.

\begin{theorem}
(Stone-Von Neuman) Up to isomorphism, there exists a unique
irreducible unitary representation $\pi : H \rightarrow
U(\mathcal{H})$ with central character $\phi$, that is,
$\pi_{|Z}=\phi\cdot Id_{\mathcal{H}}$.
\end{theorem}

The representation $\pi$ in the above theorem is
called the {\em Heisenberg representation}. In this paper, we take
a character of $Z$ as $\phi((0,0,z))=\eta^z$.
Then the unique irreducible unitary representation $\pi$
corresponding to $\phi$  has the following formula
\begin{equation}
\pi(t,w,z)[\varphi](i)=\eta^{2^{-1}tw+z+wi}\varphi(i+t)
\end{equation}
for $ \varphi\in \mathbb{C}(\mathbb{F}_p)$, $(t,w,z)\in H$.
Consequently, we have\\
$$\pi(t,0,0)[\varphi](i)=\varphi(i+t)$$
$$\pi(0,w,0)[\varphi](i)=\eta^{wi}\varphi(i)$$
$$\pi(0,0,z)[\varphi](i)=\eta^z\varphi(i).$$
Thus $\pi(t,0,0),\pi(0,w,0)$ are equal to the unitary operators
time-shift $L_t$ and phase-shift $M_w$, respectively, defined in
(\ref{eq-un}).

\subsection*{The Weil Representation}
The symplectic group $Sp = Sp(V ,\omega)$, which is isomorphic to
$SL_2(\mathbb{F}_p)$, acts by automorphism of $H$ through its action
on the $V$-coordinate, i.e., $\forall(t,w,z)\in H$ and a matrix
$g=\left(
     \begin{array}{cc}
       a & b \\
       c & d \\
     \end{array}
   \right)
\in SL_2(\mathbb{F}_p)$, the action $g$ on $(t,w,z)$ as
\begin{equation}
g\cdot (t,w,z)= (at+bw, ct+dw, z).
\end{equation}

Due to Weil \cite{A.Weil}, a projective unitary representation
$\widetilde{\rho}: Sp\rightarrow PGL(\mathcal{H}) $ is constructed
as follows. Considering the Heisenberg representation $\pi: H
\rightarrow U(\mathcal{H})$ and $g\in Sp$, a new
representation is defined as: $\pi^g: H \rightarrow U(\mathcal{H})$
by $\pi^g(h) = \pi(g(h))$. Because both $\pi$ and $\pi^g$ have the
same central character $\phi$, they are isomorphic by Theorem $4$.
By Schur's Lemma \cite{J. P. Serre}, $Hom_H$($\pi$,$\pi^g$)$\cong
\mathbb{C}^*$, so there exist a projective representation
$\widetilde{\rho}: Sp\rightarrow PGL(\mathcal{H})$. This projective
representation $\widetilde{\rho}$ is characterized by the formula:
\begin{equation}\label{eq-we}
\widetilde{\rho}(g)\pi(h)\widetilde{\rho}(g^{-1})=\pi(g(h))
\end{equation}
for every $g\in Sp$ and $h\in H$. Moreover, $\widetilde{\rho}(g)$ uniquely lifts to a unitary
representation
$$\rho: Sp\rightarrow U(\mathcal{H})$$
that satisfies equation (\ref{eq-we}). The existence of $\rho$ follows from the fact \cite{Beyl} that any
projective representation of $SL_2(\mathbb{F}_p)$ can be lifted to
an honest representation, while the uniqueness of $\rho$ follows
from the fact \cite{S.Gurevich4} that the group $SL_2(\mathbb{F}_p)$
has no non-trivial characters for $p\neq 3$.

Thus  the Weil representation,  specified in Section 4.1, follows.

\subsection*{Notion of an Unitary Representation}

Let $\mathcal{H}$ be a Hilbert space.
A unitary operator on $\mathcal{H}$ is an operator $A : \mathcal{H}\rightarrow \mathcal{H}$ which
preserves the inner product, that is, $<A\varphi, A\psi>=<\varphi, \psi>$ for every $\varphi, \psi \in \mathcal{H}$. The set of
unitary operators forms a group under composition of operators, which is denoted by
$U(\mathcal{H})$.

\begin{definition}
A {\em unitary representation} of a group $G$ on the Hilbert space $\mathcal{H}$ is a homomorphism
$\pi: G\rightarrow U(\mathcal{H})$, i.e., $\pi$ is map which satisfies the condition
$$\pi(g\cdot h)=\pi(g)\cdot \pi(h),\ \ \forall g, h \in G.$$
\end{definition}

\begin{definition}
A unitary representation $\pi: G\rightarrow U(\mathcal{H})$ is called {\em irreducible}  if there is no proper subspace $\mathcal{H}'\subset \mathcal{H}$ invariant under G, i.e., such that
$$\pi(g)\varphi \in \mathcal{H}', \ \ \forall \varphi \in \mathcal{H}'.$$
\end{definition}

All unitary representations $\pi: G\rightarrow U(\mathcal{H})$ can be decomposed
into a direct sum of irreducible unitary representations. In other words,
there exists is a decomposition of the Hilbert space $\mathcal{H}$ into a direct
sum
$$\mathcal{H}=\bigoplus_{i\in I} \mathcal{H}_i, $$
such that each subspace $\mathcal{H}_i$ is closed under the action of $G$, that is $\pi(g)\varphi \in \mathcal{H}_i,$
$\forall \varphi \in \mathcal{H}_i$, and such that the restricted unitary representations $\pi_i:G\rightarrow U(\mathcal{H}_i)$ are
irreducible.

A unitary operator $A : \mathcal{H}\rightarrow \mathcal{H}$ can be diagonalized, which means that there
exists an orthogonal decomposition of $\mathcal{H}$ into a direct sum of eigenspaces
$$\mathcal{H}=\bigoplus_{\lambda_i} \mathcal{H}_{\lambda_i}, $$
where dim($\mathcal{H}_{\lambda_i}$)$=1$, and $\forall \varphi \in \mathcal{H}_{\lambda_i}$, $A\varphi=\lambda_i \varphi$.

Now we consider a unitary representation
$\pi: G\rightarrow U(\mathcal{H})$ of a commutative group $G$, which yields a commutative group
$\{\pi(g):g\in G\}$ of unitary operators. Then the unitary operators
$\{\pi(g):g\in G\}$ can be diagonalized simultaneously,
i.e., there exists an orthogonal decomposition of $\mathcal{H}$ into common
eigenspaces
$$\mathcal{H}=\bigoplus_{i=1}^n \mathcal{H}_{\chi_i}.$$
Here the eigenspaces are indexed by the characters $\chi_i: G\rightarrow S^1$
where we have $\pi(g)\varphi=\chi_i(g)\varphi$ for every $g\in G$, $\varphi\in \mathcal{H}_{\chi_i}$.

Then we achieve the decomposition of the Weil
representation associated with maximal tori in Section 4.2.

\subsection*{PAPR and Discrete Fourier Transform}

The following theorem gives a relationship among  the continuous Fourier transform,  discrete Fourier transform and PAPR.
\begin{theorem}(\cite{Litsyn1})
Let $\varphi$ be a sequence with period $n$, and define continuous Fourier transform of $\varphi$ as
$S_{\mathbf{\varphi}}(z)=\sum_{i=0}^{n-1}\varphi(i)z^i$, then
$$\max_{|z|=1}|S_{\mathbf{\varphi}}(z)|\leqslant (\frac{2}{\pi}\ln N
+2)\max_{0\leqslant i\leqslant n-1}|F\varphi(i)|.$$
Thus $$PAPR(\varphi)\leqslant (\frac{2}{\pi}\ln N
+2)\max_{0\leqslant i\leqslant n-1}|F\varphi(i)|.$$

\end{theorem}

\end{document}